\documentclass{PoS}
\pdfoutput=1

\usepackage{graphicx}
\usepackage{amssymb}
\usepackage{amsmath}
\usepackage{xcolor}
\usepackage{listings}
\usepackage{booktabs}
\usepackage{siunitx}

\lstdefinelanguage{SVE}{ %
    keywordstyle=\bfseries\color{green!60!black},
    identifierstyle=\color{black},
    sensitive=false, 
    morecomment=[l]{//}, 
    morecomment=[s]{/*}{*/}, 
    morestring=[b]" 
}

\title{Lattice QCD on upcoming Arm architectures
}

\ShortTitle{Lattice QCD on upcoming Arm architectures}

\author{\speaker{Nils Meyer}$^{,a}$, Dirk Pleiter$^{a,b}$, Stefan Solbrig$^a$, Tilo Wettig$^a$\\
  $^a$Department of Physics, University of Regensburg, 93040 Regensburg, Germany\\
  $^b$J\"{u}lich Supercomputing Centre, Forschungszentrum J\"{u}lich, 52428 J\"{u}lich, Germany\\
  E-mail: \email{nils.meyer@ur.de}}

\abstract{ Recently Arm introduced a new instruction set called
  Scalable Vector Extension (SVE), which supports vector lengths up to
  2048 bits. While SVE hardware will not be generally available until
  about 2021, we believe that future SVE-based architectures will have
  great potential for Lattice QCD. In this contribution we discuss key
  aspects of SVE and describe how we implemented SVE in the Grid
  Lattice QCD framework. }

\FullConference{The 36th Annual International Symposium on Lattice Field Theory - LATTICE2018\\
		22-28 July, 2018\\
		Michigan State University, East Lansing, Michigan, USA.}

\begin{document}

\newbox\bothboxb{}
\setbox\bothboxb=\hbox{
\begin{lstlisting}[numbers=left]
  void daxpy(double a, double *restrict x, double *restrict y, int n) {
    for (int i = 0; i < n; i++)
      y[i] = y[i] + a * x[i];
  }
\end{lstlisting}
}

\section{Introduction}


Arm is a vendor of intellectual property (IP).
Processors based on Arm IP are used in many kinds of computing devices.
For instance, Arm is the most popular architecture for mobile processors.
Silicon providers
also started to address the server market, e.g., Marvell implements
Arm IP in their ThunderX2 system-on-a-chip (SoC) processor.  ThunderX2 cores support
128-bit SIMD instructions known as NEON.
The SoCs ship in various configurations featuring up to 32 of such cores clocked
at a maximum frequency of 2.5 GHz.  They support 8 channels of DDR4 memory,
i.e., the nominal memory bandwidth is higher compared to current Intel Xeon processors.

The Isambard project \cite{Isambard} is the first HPC production system leveraging
Arm technology.  It is run as a Tier 2 service in the UK.  Isambard is a Cray
XC50 architecture that consists of 328 ThunderX2 servers interconnected by
a Cray Aries interconnect.
Another vendor of Arm-based HPC is Hewlett Packard Enterprises (HPE).
Astra at Sandia National Labs, an HPE machine, is the first Petascale system based
on Arm technology and achieved rank 204 in the Top500 list of November
2018.  The architecture consists of 5184 ThunderX2 servers interconnected by
InfiniBand.

In 2016 Arm announced a novel instruction set called Scalable Vector Extension (SVE).
SVE comprises vector instructions with input operands of length 128 to 2048 bits.
The Post-K supercomputer, successor to the Japanese flagship K-Computer,
is scheduled to start operation at RIKEN (Kobe) in 2021.  The Post-K incorporates Fujitsu
A64FX processors interconnected by a custom-designed toroidal network.
The A64FX processor is specifically designed for the Post-K and expected to
be the first hardware implementation to support SVE, in this case a 512-bit version.

We consider SVE attractive for applications in Lattice QCD.  In the framework
of the QPACE~4 project
we pursue evaluation and enhancement of the existing SVE software and upcoming
SVE hardware technologies.

In this contribution we discuss key aspects of the SVE instruction set and
programming model.  We give an overview of the architecture of
the Fujitsu A64FX processor and comment on how we support the NEON
and SVE instruction sets in the Lattice QCD framework Grid
\cite{Boyle:2015tjk}.

\section{Arm Scalable Vector Extension}

\subsection{Overview}

SVE is a vector extension for Arm
architectures.  Here we list only the most relevant features for Lattice QCD
applications and refer to \cite{DBLP:journals/corr/abs-1803-06185} for a comprehensive overview.
\begin{itemize}
  \item Variable-length vector units with a length of 128 to 2048 bits
        (the length must be a multiple of 128 bits);
  \item Vectorized 16-, 32- and 64-bit floating-point operations, including
  fused multiply-add and conversion of precision;
  \item Vectorized arithmetic of complex numbers, including fused multiply-add.
\end{itemize}
The instruction set is identical for all
implementations. It is up to the hardware vendor to decide on a specific vector length
for a given target architecture.

\subsection{Upcoming hardware}

\begin{figure}
	\centering
  \includegraphics[width=0.8\textwidth]{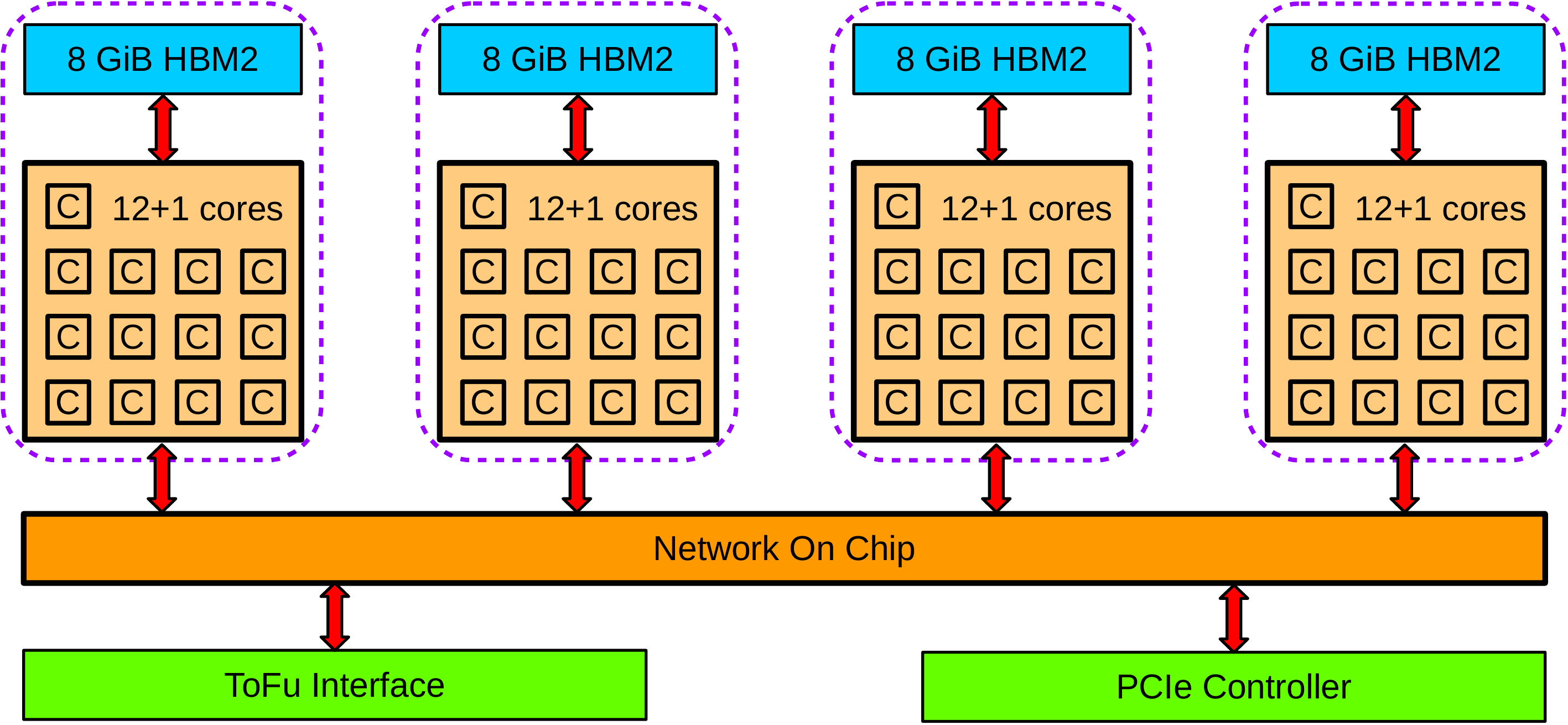}
	\caption{High-level architecture of the Fujitsu A64FX processor
  \cite{Fujitsu:a64fx}.}
	\label{fig:a64fx}
\end{figure}

SVE hardware is not available yet.  The Fujitsu A64FX processor mentioned in the introduction is the first,
and at present the only, SVE implementation publicly announced
\cite{Fujitsu:a64fx}.

The high-level architecture of the A64FX is shown in Fig.~\ref{fig:a64fx}.  The CPU
contains 48 + 4 identical out-of-order cores with 512-bit vector length
and is connected to \SI{32}{GiB} of in-package main memory.
Cores and main memory are arranged as four cache-coherent NUMA nodes called Core
Memory Groups (CMG).  Each CMG consists of 12 compute cores and 1
assistant core, \SI{8}{MiB} shared L2 cache and \SI{8}{GiB} HBM2 high-bandwidth memory.
The peak aggregate main memory throughput of the chip is \SI{1024}{GiB/s}.  The CMGs
are interconnected by a network-on-a-chip (NoC), implemented as a ring bus. The NoC
connects to the PCIe GEN3 x16 interface and the proprietary Torus Fusion
(ToFu) network interface.

The A64FX executes up to two fused multiply-add instructions per core and per cycle.
The chip achieves an aggregate peak compute performance of at least \SI{2.7}{TFlop/s}
in double precision.  At present there is no information about the performance
characteristics of the full instruction set and the CPU frequency available.
Given the performance data, we estimate the CPU frequency to be at least \SI{1.7}{GHz}.

\subsection{SVE features}
\label{sec:vla}

A key feature of SVE is that it is Vector-length Agnostic (VLA). This means that
the vector length is unknown at compile time.
Code execution adapts to the hardware implementation with a specific vector length
at runtime.
Hardware features that enable VLA include increment instructions, which increment a variable
depending on the actual vector length, and support of predication, i.e., the ability to
execute operations only on parts of the input vectors depending on the content of a
predication register.

As an illustrative example for the implications of VLA we discuss the \texttt{daxpy}
kernel, which evaluates $y_i \leftarrow y_i + a\times x_i$ with real operands
in double precision.  The implementation in C99 is shown in
Fig.~\ref{fig:daxpyc99}.

\begin{figure}
\begin{lstlisting}[language=C++, numbers=left, xleftmargin=0.05\textwidth]
void daxpy(double a, double *restrict x, double *restrict y, int n) {
    for (int i = 0; i < n; i++)
        y[i] = y[i] + a * x[i];
}
\end{lstlisting}
\caption{Implementation of the \texttt{daxpy} kernel in C99.}
\label{fig:daxpyc99}
\end{figure}

We use the auto-vectorization capability of the armclang SVE compiler.
armclang is unaware of the vector length and optimizes following the
VLA paradigm.  The assembly output is shown in Fig.~\ref{fig:daxpyasm}.

\begin{figure}
\begin{lstlisting}[language=SVE, numbers=left, xleftmargin=0.05\textwidth]
    mov      x9, xzr                          // xzr=0 => x9=0
    whilelo  p1.d, xzr, x8                    // predication for x_{i...}, y_{i...}
                                              // x8=n; z_0=a
    ptrue    p0.d                             // predication for result vector
.LBB0_2:
    ld1d     {z1.d}, p1/z, [x1, x9, lsl #3]   // predicated load z1 <- y_{i...}
    ld1d     {z2.d}, p1/z, [x2, x9, lsl #3]   // predicated load z2 <- x_{i...}
    fmad     z1.d, p0/m, z0.d, z2.d           // predicated multiply-add
    st1d     {z1.d}, p1, [x2, x9, lsl #3]     // predicated store z1 -> y_{i...}
    incd     x9                               // increment index x9 by vector length
    whilelo  p1.d, x9, x8                     // predication for remaining y_{i...}
    b.mi     .LBB0_2                          // conditional branch to LBB0_2
\end{lstlisting}
\caption{SVE assembly of the \texttt{daxpy} kernel generated by the armclang SVE
compiler.}
\label{fig:daxpyasm}
\end{figure}

We restrict ourselves to commenting on key aspects of the assembly
and refer to \cite{Meyer:cluster18} for a detailed discussion.
The compiler creates a single loop containing vector instructions.
The SVE vector length is unknown at compile time.
Using predication, the operands are
loaded from memory into registers, processed, and the result array is stored
in memory.  Predication masks the vector lanes valid for processing.
Furthermore, predication eliminates the need for front-end and/or tail
recursion, which is required for typical fixed-size SIMD if the length of the
array is not a multiple of the SIMD width.
Note that the SVE vector length does not appear explicitly.
The increment of the array index takes
this length into account implicitly.  The number of loop
iterations is thus determined at runtime.  The code executes on architectures
with different SVE vector lengths without the need for recompilation.

It is not guaranteed that the compiler exploits all features of
the SVE instruction set. For instance, we are using the armclang compiler, which currently does
not support vector instructions for complex arithmetics.
This feature of SVE can nevertheless be exploited
using so-called intrinsic functions. These are functions that are treated by the compiler
in a special way and usually do not result in function calls.
The Arm C Language Extension (ACLE)
defines intrinsics for SVE \cite{ARM:ACLE}.\footnote{ACLE intrinsics are also
available for NEON.}  SVE ACLE also defines vector data types which reflect the VLA
paradigm: vector data types do not have a defined size.  They are referred to as
''sizeless structs.''  Due to their sizeless nature they must not be used
\begin{itemize}
  \item as data members of unions, structures and classes;
  \item to declare or define a static or thread-local storage variable; or
  \item as the argument of \texttt{sizeof}.
\end{itemize}
These restrictions have impact on the strategy porting our target Lattice QCD
framework Grid to SVE, which we comment on in Section \ref{sec:grid} and discuss
in detail in \cite{Meyer:cluster18}.

\subsection{SVE code development}

While NEON hardware exists and the corresponding toolchain, including
clang/LLVM, gcc and gdb, is well established, the situation is very
different for SVE. No SVE hardware exists, and the tool support is
still in development.

We use engineering samples of the Cavium ThunderX2 for code development.
Arm provides us with their proprietary LLVM/clang-based armclang SVE compiler
\cite{ARM:hpctools}
in the context of a research contract.  armclang supports ACLE intrinsics and VLA.
It is unaware of the vector length and does not support fixed-size
SVE.  We use armclang for compiling and gdb for debugging our codes.
gdb officially supports SVE as of version 8.2.  For the future we plan to
evaluate gcc 9, which is announced for 2019 featuring SVE intrinsics support,
and also Fujitsu's SVE compiler.

We use the Arm Instruction Emulator (ArmIE) for emulation of SVE binaries
\cite{ARM:hpctools}.  The vector length is set as a command-line parameter.  We
contributed to ArmIE extending its functionalities for code instrumentation.
ArmIE requires native 64-bit Arm hardware.

As emulation of SVE does not allow for a performance evaluation, we also started
to simulate architectures supporting SVE using the gem5 simulator.
gem5 is a modular platform for computer architecture research, encompassing
system-level and processor architecture \cite{gem5}.  RIKEN provides us
with access to their gem5 simulator of the Fujitsu A64FX processor.  We will
optimize our code guided by simulations of the A64FX.

\section{Lattice QCD on Arm architectures}
\label{sec:grid}

Grid \cite{Boyle:2015tjk} is a highly portable Lattice QCD framework
with support for a multitude of parallel architectures.  Currently Grid supports
vector lengths up to 512 bits.  Recently efforts have been made porting Grid to
nVidia GPGPUs.

In this contribution we focus on the CPU implementation.  The portability issue
is resolved by implementation of architecture-specific low-level
functions using intrinsics, e.g., for arithmetics of complex
numbers, thereby achieving 100\% SIMD-efficiency.\footnote{Performance-critical
kernels such as the Wilson-Dirac operator are implemented in assembly for several
architectures.}  The core of Grid's abstraction layer is a template class that
enables direct access to vector registers using intrinsics data types. These
data types are declared as member data.  The implementation scheme is feasible
due to the compilers' capability to auto-generate loads/stores addressing
intrinsics data types.

To illustrate the intrinsics implementation for fixed-size SIMD we show the
multiplication of complex numbers for 128-bit Arm NEON architectures, which we
contributed to Grid in 2017, in Fig.~\ref{fig:gridneon}.  Grid stores complex
numbers within two-element structures composed of real and imaginary part.
Arithmetics of real operands and permutations of vector elements are necessary
for complex multiplication.

For enabling SVE in Grid we have to respect the restrictions on usage of SVE
ACLE intrinsics data types pointed out in Section~\ref{sec:vla}.  We describe our strategy for adapting Grid to SVE in
detail in \cite{Meyer:cluster18}.
Key aspects of our SVE-enabled implementation are the following:
\begin{itemize}
\item We fix the vector length at compile time, thereby mitigating restrictions
on vector data types and omitting unnecessary loops implied by VLA.
\item The variety of predications is minimized to fit into the register file.
\item Hardware support for relevant features of the SVE instruction set, e.g.,
complex arithmetics, is added using intrinsics.
\end{itemize}
It is not a priori clear which code implementation achieves the best
performance.  Therefore, we implemented multiple implementation options, e.g., for
processing complex numbers.  The source code is available at
\cite{Meyer:gridgithub}.  We illustrate one implementation option for
multiplication of complex numbers in Fig.~\ref{fig:gridsve}.
\begin{figure}[t]
\begin{lstlisting}[language=C++, numbers=left, xleftmargin=0.05\textwidth]
struct MultComplex {
    inline float64x2_t operator()(const float64x2_t &a, const float64x2_t &b) {
        float64x2_t r0, r1, r2, r3, r4;
        r0 = vtrn1q_f64(b, b);
        r1 = vnegq_f64(b);
        r2 = vtrn2q_f64(b, r1);
        r3 = vmulq_f64(r2, a);
        r4 = vextq_f64(r3, r3, 1);
        return vfmaq_f64(r4, r0, a);
    }
};
\end{lstlisting}
\caption{NEON ACLE implementation of vectorized multiplication of complex
numbers in double precision.  Permutations of vector elements are necessary
prior to computation using real arithmetics.  Load/store operations are
auto-generated by the compiler.}
\label{fig:gridneon}
\end{figure}

\begin{figure}
\begin{lstlisting}[language=C++, numbers=left, xleftmargin=0.05\textwidth]
template <typename T>
struct vec {
    alignas(VL) T v[VL / sizeof(T)];
};

struct MultComplex{
    template <typename T>
    inline vec<T> operator()(const vec<T> &x, const vec<T> &y) {
        vec<T> out;
        svbool_t pg1 = acle<T>::pg1();
        typename acle<T>::vt z_v = acle<T>::zero();
        typename acle<T>::vt x_v = svld1(pg1, x.v);
        typename acle<T>::vt y_v = svld1(pg1, y.v);
        typename acle<T>::vt r_v = svcmla_x(pg1, z_v, x_v, y_v, 90);
        r_v = svcmla_x(pg1, r_v, x_v, y_v, 0);
        svst1(pg1, out.v, r_v);
        return out;
    }
};
\end{lstlisting}
\caption{SVE ACLE implementation of vectorized multiplication of complex
numbers in single and double precision. The vector length (VL) is fixed at
compile time.  The \texttt{svcmla} intrinsic addresses hardware support for
complex multiplication without the need for permutations of the vector elements,
cf.~\cite{Meyer:cluster18, ARM:ACLE}.
Explicit load/store operations copy the data to/from vector registers. The
template class \texttt{acle<T>} simplifies the mapping of C++ and SVE ACLE data types
and provides various definitions of predication.}
\label{fig:gridsve}
\end{figure}

\section{Summary and outlook}

We introduced the Arm Scalable Vector Extension (SVE), which we
believe will attain significant attention in the HPC community once SVE hardware
is available.  In the framework of the QPACE~4 project we contribute to
enhancements of the SVE toolchain and port the Grid Lattice QCD framework
to SVE.  For the future we plan to evaluate other SVE compilers and to
optimize our code using RIKEN's gem5 simulator of the Fujitsu A64FX processor.

\section*{Acknowledgment}

We acknowledge funding of the QPACE~4 project provided by the Deutsche
Forschungsgemeinschaft (DFG) in the framework of SFB/TRR-55.  Furthermore, we
acknowledge support from the HPC tools team at Arm and from the
Post-K team at RIKEN.  We thank Peter Boyle, Guido Cossu and Mitsuhisa Sato for valuable
discussions.

\bibliographystyle{JHEP_lat18}
\bibliography{references}

\providecommand{\href}[2]{#2}\begingroup\raggedright\begin{thebibliography}{1}

\bibitem{Isambard}
Arm, {\it {Case study: GW4 Alliance puts Arm architecture to the test for
  HPC}},
  [\href{https://developer.arm.com/-/media/developer/products/software-tools/hpc/White\%20papers/Isambard_GW4_Whitepaper_Digital.pdf}{https://developer.arm.com/-/media/developer/products/software-tools/hpc/White\%20papers/Isambard\_GW4\_Whitepaper\_Digital.pdf}].

\bibitem{Boyle:2015tjk}
P.~Boyle, A.~Yamaguchi, G.~Cossu, and A.~Portelli, {\it {Grid: A next
  generation data parallel C++ QCD
  library,}}\href{http://dx.doi.org/10.22323/1.251.0023}{ {\em PoS (LATTICE
  2015)} 023} [\href{http://arxiv.org/abs/1512.03487}{{\tt arXiv:1512.03487}}].

\bibitem{DBLP:journals/corr/abs-1803-06185}
N.~Stephens et~al., {\it {The {ARM} Scalable Vector
  Extension,}}\href{http://dx.doi.org/10.1109/MM.2017.35}{ {\em IEEE Micro}
  {\bf 37} (2017) 26} [\href{http://arxiv.org/abs/1803.06185}{{\tt
  arXiv:1803.06185}}].

\bibitem{Fujitsu:a64fx}
T.~Yoshida, {\it {Fujitsu High Performance CPU for the Post-K Computer}}, Hot
  Chips 2018
  [\href{http://www.fujitsu.com/jp/Images/20180821hotchips30.pdf}{http://www.fujitsu.com/jp/Images/20180821hotchips30.pdf}].

\bibitem{Meyer:cluster18}
N.~Meyer, P.~Georg, D.~Pleiter, S.~Solbrig, and T.~Wettig, {\it {SVE-enabling
  Lattice QCD Codes,}}\href{http://dx.doi.org/10.1109/CLUSTER.2018.00079}{ {\em
  {2018 IEEE International Conference on Cluster Computing (CLUSTER)}} 623}
  [\href{http://arxiv.org/abs/1901.07294}{{\tt arXiv:1901.07294}}].

\bibitem{ARM:ACLE}
Arm, {\it {ARM C Language Extension for SVE}},
  [\href{https://developer.arm.com/docs/100987/latest/arm-c-language-extensions-for-sve}{https://developer.arm.com/docs/100987/latest/arm-c-language-extensions-for-sve}].

\bibitem{ARM:hpctools}
Arm, {\it {ARM HPC tools}},
  [\href{https://developer.arm.com/products/software-development-tools/hpc/documentation}{https://developer.arm.com/products/software-development-tools/hpc/documentation}].

\bibitem{gem5}
gem5, [\href{http://gem5.org/Main_Page}{http://gem5.org/Main\_Page}].

\bibitem{Meyer:gridgithub}
N.~Meyer et~al., {\it {SVE-enabled Grid}},
  [\href{https://github.com/nmeyer-ur/Grid/tree/feature/arm-sve}{https://github.com/nmeyer-ur/Grid/tree/feature/arm-sve}].

\end{thebibliography}\endgroup

\end{document}